\documentclass[12pt]{iopart}

\begin{document}

\title[Light absorption in deformed graphene]{Light absorption in deformed graphene}

\author{Sa\'ul Hern\'andez-Ortiz$^1$, David Valenzuela$^2$,  Alfredo Raya$^{1}$, Sa\'ul S\'anchez-Madrigal$^{3}$}

\address{$^1$Instituto de F\'{\i}sica y Matem\'aticas, Universidad Michoacana de San Nicol\'as de Hidalgo, 
Edificio C-3, Ciudad Universitaria, 58040 Morelia, Michoac\'an, M\'exico.\\
$^2$Instituto de F\'{\i}sica, Pontificia Universidad Cat\'olica de Chile, 
Casilla 306, Santiago 22, Chile.\\
$^3$Facultad de Ciencias, Universidad de Colima, Bernal D\'{\i}az del Castillo 340, Colima, Colima 28045, M\'exico.}

\ead{sortiz@ifm.umich.mx, devalenz@uc.cl, raya@ifm.umich.mx, saul\_sanchez@ucol.mx}

\begin{abstract}
We model the low energy dynamics of graphene in the continuum in terms of a version of Reduced Quantum Electrodynamics restricting fermions  to a (2+1)-dimensional brane, whilst photons remain within the (3+1)-dimensional bulk. For charge carriers, besides the Dirac mass gap, we consider a Haldane mass term which is induced by parametrizing an effective parity $\mathcal{P}$ and time-reversal  $\mathcal{T}$ symmetry breaking that occurs on the brane when deformations of the honeycomb array are such that the equivalence between sublattices  is lost. We make use of the relativistic Kubo formula and carry out an explicit calculation of the transverse conductivity. As expected, the filling factor is a half (in natural units) for each fermion species. Furthermore, assuming that a sample of this material is radiated perpendicularly with polarized monochromatic light of frequency $\omega$, from the modified Maxwell's equations we study the problem of light absorption in graphene in terms of the said conductivity. We observe an analog effect to the Faraday Rotation due the Induced Mass (FRIM)--and not to an external magnetic field-- in which light penetrating the sample changes its angle of polarization solely by effect of the induced mass. This effect might be relevant for the development of optic filters based on mechanical stretching of graphene flakes. 
\end{abstract}

\section{Introduction}	

Graphene is a two dimensional  array of carbon atoms in a hexagonal lattice. It is one to the most interesting materials in solid state physics nowadays due to its  exceptional properties and its potential nanothechnological applications, one of the reasons perhaps for which Andrei Geim and Konstantin Novoselov were Laureated with a Nobel Prize in Physics in 2010. Progress in the experimental isolation of monolayer graphene samples~\cite{graphene1,graphene2,graphene3} has led to an extensive exploration of the electronic properties of this material. Its crystal structure allows an accurate tight-binding description at low energies, which becomes, in the continuous limit, a version of massless quantum electrodynamics in (2+1)-dimensions, QED$_3$, for the charge carriers restricted to move along a membrane~\cite{Gusynin:review}, but in which photon are allowed to move throughout space in such a way that the static Coulomb interaction is still described by a potential that varies as the inverse of the distance on the plane of motion of quasiparticles. This model incorporates the most essential and well-established properties of the charge carrier dynamics: the symmetries of the hexagonal lattice, the linearity of the dispersion relation, a very small mass gap (if any), and a characteristic propagation velocity which is 1/300 of the speed of light in vacuum~\cite{Gusynin:review,review2,properties}.

A remarkable feature of graphene is the visual transparency of its membranes. Opacity of layers of this crystal has been measured~\cite{measure} to be roughly 2.3\% with almost negligible reflectance. This observation has opened the possibility of using monolayer graphene in combination with bio-materials to produce clean hydrogen by photocatalysis~\cite{hydrogen} with visible light. The problem of light absorption in graphene can be addressed from quantum field theoretical arguments~\cite{fial1,fial2,fial3,Nosotros}, considering  the Dirac picture for its charge carriers  in terms of the degrees of freedom of  QED$_3$ under different assumptions.  Parity violating effects were considered in~\cite{fial2}, whereas the influence of a strong and weak magnetic field were considered in~\cite{fial3} and~\cite{Nosotros}, respectively. Measurements of magneto-optical properties of epitaxial graphene have been reported in Ref.~\cite{faradayexp}, particularly the polarization rotation and light absorption. Quantum Faraday and Kerr rotations have also been experimentally determined~\cite{kerr} and a complete framework based on the equations of motion was presented in~\cite{aires} to describe those effects.

In this article we explore an analog to the Faraday effect in deformed graphene. We consider deformations of the membrane which break parity ${\cal P}$ and time-reversal ${\cal T}$ symmetries, rendering the sublattices inequivalent. Examples of such deformations are, for instance, strains~\cite{strain}, known to give rise to homogeneous pseudomagnetic fields. The net effect of deformations is parametrized in terms of a ${\cal P}$ and ${\cal T}$ breaking Haldane mass~\cite{Haldane} which  
plays the role of an external magnetic field. 
We find that the mentioned mass term induces a change in the angle of polarization for monochromatic light passing perpendicularly through the graphene sample. Our key observation is that the Haldane mass induces a half-filling Quantum Hall Effect (QHE) per fermion species (see, for instance,~\cite{Edward}), and because the change of the angle of polarization of light passing through a graphene membrane can be expressed entirely in terms of the Hall conductivity~\cite{fial1,fial2,fial3,Nosotros}, the effect we propose in this work can be expressed solely in terms of the mass parameter that accounts for deformations in graphene. The effect we observe can be the basis of an optical filter based on mechanical deformation of graphene. To describe our findings, we have organized the remaining of this article as follows: In Sec. 2 we introduce a continuous model for the charge carriers in graphene and explicitly show the decomposition of the propagator for two different fermions species according to their {\em chiralities}.  We provide a proper meaning of this property in terms of the lattice structure of graphene. In Sec. 3 we calculate the conductivity tensor through the filling factor using the Kubo formula and in Sec. 4 we review the problem of light absorption in graphene through the modified Maxwell's equations which describe the penetration of electromagnetic waves into the sample. From the matching conditions, we calculate the transmission coefficients and  the  angle of polarization rotation of transmitted light. Discussions and conclusions are presented in Sect. 5.

\section{A continuous model for graphene}

Tight-binding approach to the description of monolayer graphene corresponds in the continuum to a massless version of quantum electrodynamics in (2+1) dimensions, QED3, but with a static Coulomb interaction which varies as the inverse of the distance, just as in ordinary space~\cite{Gusynin:review}. The dynamics can be modeled from the action
\begin{equation}
I[\psi,A]=-\frac{1}{4}\int d^4x F_{\hat{\mu}\hat{\nu}}^2+\int d^3x \bar\psi {\not \! D}\psi\;,
\label{action}
\end{equation}
with $F_{\hat{\mu}\hat{\nu}}=\partial_{\hat{\mu}} A_{\hat{\nu}}-\partial_{\hat{\nu}} A_{\hat{\mu}}$, ${\not\!D}=i\tilde{\gamma}^\mu(\partial_\mu+ieA_\mu)$, $A_{\hat{\mu}}$ representing the gauge field that propagates in (3+1)-dimensions and $\psi$ the quasiparticle fermion field. The symmetry which exists between the two triangular sublattices allows to merge the fermion content into a single four-component theory for the charge carriers. In our considerations, circumflexed greek indices  $\hat{\alpha},~\hat{\beta},~\hat{\gamma}$ and so on take the values 0, 1, 2, 3; greek indices  $\mu,~\nu,~\gamma,$ etcetera run from 0 to 2;  and latin indices $a,~b$ and so on --which label the spatial coordinates on the graphene layer-- take the values 1 and 2. Moreover, the re-scaled $4\times4$ Dirac matrices are such that $\tilde{\gamma}^0=\gamma^0$, $\tilde{\gamma}^{1,2}=v_F\gamma^{1,2}$ and for later convenience, we also consider the matrices $\tilde{\gamma}^3=\gamma^3$ and $\tilde{\gamma}^5=\gamma^5$, where $v_F$ is the Fermi velocity of quasiparticles in the crystal. In the natural units of the system (namely, when $v_F=1$), the form of the Lagrangian has been dubbed as Reduced QED and has been proposed in the context of brane-world scenarios~\cite{RQED}. Explicitly, we consider
\begin{equation*}
\tilde{\gamma}^0=\left(
\begin{array}{cc}
\sigma_3 & 0 \\
0 & -\sigma_3
\end{array}
\right),\quad \tilde{\gamma}^i=v_F\left(
\begin{array}{cc}
i\sigma_i & 0 \\
0 & -i\sigma_i
\end{array}\right) \;, 
\end{equation*}
\begin{equation}
\tilde{\gamma}^3=\left(
\begin{array}{cc}
0 & I \\
I & 0
\end{array}
\right),\quad \tilde{\gamma}^5=i\left(
\begin{array}{cc}
0 & I \\
-I & 0
\end{array}\right) \; .\label{Gamma}
\end{equation}
Notice that in this case, the free, massless Dirac Lagrangian 
\begin{equation}
\mathcal{L}=\bar{\psi}\ i{\not\! \partial}\ \psi \;,\label{massless}
\end{equation}
is invariant under the two chiral-like transformations 
\begin{equation}
\psi \to e^{i\sigma \tilde{\gamma}^3\psi}\;, \qquad \mbox{and} \qquad \psi \to e^{i\beta \tilde{\gamma}^5\psi} \; ,\label{chiral}
\end{equation}
which means that besides the ordinary Dirac mass term, we can introduce a second term $m_0\bar\psi \tau \psi$ with $\tau=i[\tilde{\gamma}^3,\tilde{\gamma}^5]/2$, referred to as the Haldane mass term~\cite{Haldane}. Such a term is invariant under the transformations in Eq.~(\ref{chiral}). However, it breaks parity $\mathcal{P}$ and time-reversal $\mathcal{T}$ symmetries. We stress that the chiral character of the transformations~(\ref{chiral}) is not related to the quasiparticle spin, but rather to their pseudospin. The massive Lagrangian thus becomes
\begin{equation}
\mathcal{L}=\bar{\psi}\left(i{\not\! \partial}-m_e-m_0\tau\right)\psi \; .\label{lag1}
\end{equation}
Introducing the chiral-like projection operators $\chi_\pm=\frac{1}{2}\left(1\pm\tau\right)$ which verify $\chi_{+}+\chi_-=1,$ and $\chi_\pm^2=\chi_\pm,$ we can re-express the Lagrangian~(\ref{lag1}) in the form
\begin{equation}
\mathcal{L}=\bar{\psi}^+\left(i{\not\! \partial}-m_+\right)\psi^++\bar{\psi}^-\left(i{\not\!\partial}-m_-\right)\psi^- \; ,\label{lag2}
\end{equation}
where $m_\pm=m_e\pm m_0$ and the ``left-'' and ``right-handed'' quasiparticle fields are $\psi^\pm=\chi_\pm \psi$. Hence, the Lagrangian describes two different charge carrier species which are non degenerated in mass: one of them becomes lighter and the other heavier as $m_0$ increases. This difference implies that the interaction between the two species in the reciprocal lattice is different. This type of asymmetry that can be introduced as a constant deformation on the graphene membrane that modifies the fundamental cell, but not the periodicity of the sample, which means that as one deforms the graphene membrane along one direction, $m_0$ grows and in consequence, the differences between the species is more evident. 

The decomposition of the free quasiparticle propagator corresponding to~(\ref{lag2}) is
\begin{eqnarray}
S(p)&=&-\left(\left[\frac{{\not\! p} + m_+}{p^2-m_{+}^2}\right]\chi_+ + \left[\frac{{\not\! p} + m_-}{p^2-m_{-}^2}\right]\chi_-\right)\nonumber\\
&=&-\left(S_+(p)\chi_+ + S_-(p)\chi_-\right) \; .\label{Schiral}
\end{eqnarray}
We shall use the above expressions for the fermion propagator to discuss the problem of the Faraday rotation due the induced mass (FRIM) in graphene.

\section{Filling Factor}

It is well known that the Hall conductivity for the QHE can be expressed as
\begin{equation}
\sigma = \left(
\begin{array}{cc}
0 & -\nu\frac{e^2}{2\pi} \\ 
\nu\frac{e^2}{2\pi} & 0
\end{array} \right)\; ,\label{SigmaFilling}
\end{equation}
where $\nu$ is a small integer (integer QHE), or a fraction (fractional QHE), the so-called filling factor. In the absence of magnetic fields, we can still have QHE provided parity and/or time reversal symmetries are broken in the Lagrangian and one form to achieve this is to consider the Haldane mass term~\cite{Edward}. A helpful tool for the calculation of the transverse conductivity via the filling factor is the Kubo formula, which expresses the linear response of an observable quantity due to a time-dependent perturbation. In terms of the fermion propagator, the filling factor is found through~\cite{Edward,Swamy} 
\begin{equation}
\nu= \frac{1}{24\pi^2}\int d^3p \epsilon_{\mu\nu\rho} \mathrm{Tr}[(\partial_\mu S^{-1})S(\partial_\nu S^{-1})S(\partial_\rho S^{-1})S]\; ,
\end{equation}
where $S$ is the electron propagator, $\epsilon_{\mu\nu\rho}$ the Levi-Civita Symbol and $\partial_\mu=\partial/\partial p^\mu$. Using the chiral-like decomposition of Eq.~(\ref{Schiral}) for the free quasiparticle propagator, the right- and left-handed projections of the filling factor can be obtained as~\cite{Edward}
\begin{equation}
\nu_{\pm}= \frac{1}{24\pi^2}\int d^3p \epsilon_{\mu\nu\rho} \mathrm{Tr}[(\partial_\mu S_{\pm}^{-1})S_{\pm}(\partial_\nu S_{\pm}^{-1})S_{\pm}(\partial_\rho S_{\pm}^{-1})S_{\pm}\chi_\pm],
\end{equation}
which shows the helpfulness of the chiral-like projectors. Hence, the filling factor is
\begin{equation}
\nu=\nu_+ +\nu_-\; .\label{FillingSum}
\end{equation} 
The  computation of $\nu_{\pm}$ reduces to calculate a trace 
\begin{equation}
\textrm{Tr}[{\cal S}_\pm]=\textrm{Tr}\left[\gamma^\mu\left[\frac{{\not\! p} + m_{\pm}}{p^2-m_{\pm}^2}\right]\gamma^\nu\left[\frac{{\not\! p} + m_{\pm}}{p^2-m_{\pm}^2}\right]\gamma^\rho\left[\frac{{\not\! p} + m_{\pm}}{p^2-m_{\pm}^2}\right]\chi_\pm\right] \; .\label{Trpm}
\end{equation}
For the representation in Eq~(\ref{Gamma}), the traces of the gamma matrices with the $\chi_\pm$ projectors satisfy:
\begin{eqnarray}
\textrm{Tr}[\chi_{\pm}]= 2 \;,\nonumber &\\
\textrm{Tr}[\gamma^{\mu}\chi_{\pm}]= 0 \;,\nonumber &\\
\textrm{Tr}[\gamma^{\mu}\gamma^{\nu}\chi_{\pm}]= 2g^{\mu\nu} \;,\nonumber &\\
\textrm{Tr}[\gamma^{\mu}\gamma^{\nu}\gamma^{\alpha}\chi_{\pm}]= \mp 2i\epsilon^{\mu\nu\alpha} \;,\nonumber &\\
\textrm{Tr}[\gamma^{\mu}\gamma^{\nu}\gamma^{\alpha}\gamma^{\beta}\chi_{\pm}]= 2\left(g^{\mu\nu}g^{\alpha\beta}-g^{\mu\alpha}g^{\nu\beta}+g^{\mu\beta}g^{\nu\alpha}\right) \;.&
\end{eqnarray} 
Notice that only $\textrm{Tr}[\gamma^{\mu}\gamma^{\nu}\gamma^{\alpha}\chi_{\pm}]$ is asymmetric upon contraction with  $\epsilon_{\mu\nu\rho}$, and thus the trace in Eq.~(\ref{Trpm}) reduces simply to
\begin{equation}
\epsilon_{\mu\nu\rho}\textrm{Tr}[{\cal S}_\pm] = i\frac{\pm 12 m_{\pm} p^2 \mp 12m^3_{\pm}} {\left(p^2 - m^2_{\pm}\right)^3}\;.
\end{equation}
Then, replacing into the Kubo formula and after Wick rotating to Euclidean space, we obtain that
\begin{equation*}
\nu_{\pm}= \mp \frac{m_\pm}{2\pi^2}\int \frac{d^3p}{\left(p^2 - m^2_{\pm}\right)^2}= \mp\frac{m_\pm}{2\pi^2}\frac{\pi^2}{|m_\pm|}=\mp\frac{1}{2}\frac{m_\pm}{|m_\pm|},
\end{equation*}
which for $m_\pm \rightarrow 0$ becomes
\begin{equation}
\nu_{\pm}=\mp\frac{1}{2}\textrm{sign}(m_\pm).\label{nuhalf}
\end{equation}
Notice that the above result~(\ref{nuhalf}) also arises naturally when considering a tight-biding model for next-to-nearest neighbors; in the continuum, the corresponding effective Hamiltionian can be straightforwardly diagonalized within a non-conmutative quantum mechanics framework~\cite{marcelo}.
Accordingly, from Eq.~(\ref{FillingSum}) the filling factor is
\begin{equation}
\nu=-\frac{1}{2}\textrm{sign}(m_+)+\frac{1}{2}\textrm{sign}(m_-)\; ,
\end{equation}
that gives $\nu=0$ if $m_0=0$, and $\nu=-1$ if $m_e=0$. In the latter case, the Hall conductivity becomes
\begin{equation}
 \sigma_{xy}=-\nu\frac{e^2}{2\pi}\;,
\label{sigmaxy}
\end{equation}
which becomes the key ingredient to address the problem of light absorption in graphene below.

\section{Light absorption}

From the action in Eq.~(\ref{action}), we can describe the propagation of electromagnetic waves through the graphene sample according to the modified Maxwell's equations
\begin{equation}
\partial_{\hat{\mu}} F^{\hat{\mu}\hat{\nu}}+\delta(z) \Pi^{\hat{\nu}\hat{\rho}}A_{\hat{\rho}}=0\;,\label{maxwell}
\end{equation}
subject to the conditions
\begin{eqnarray}
A_{\hat{\mu}}\Bigg|_{z=0^+}-A_{\hat{\mu}}\Bigg|_{z=0^-}&=&0\;,\nonumber\\
(\partial_z A_{\hat{\mu}})\Bigg|_{z=0^+}-(\partial_z A_{\hat{\mu}})\Bigg|_{z=0^-}&=&\Pi_{\hat{\mu}}^{\hat{\nu}} A_{\hat{\nu}} \Bigg|_{z=0}\;.
\label{bc}
\end{eqnarray}
In the above expressions, $\Pi^{\hat{\nu}\hat{\rho}}$ represents the vacuum polarization tensor, which has the general form
\begin{equation}
\Pi^{\mu \nu} = \alpha \left[ \Psi(p)\left(g^{\mu \nu}- \frac{p^\mu p^\nu}{p^2}\right)+i\Phi(p)\epsilon^{\mu\nu\rho} p_\rho\right]\;,
\end{equation}
$\Psi(p)$ and $\Phi(p)$ representing polarization scalar functions. The first term describes the part that preserves parity and has been used to estimate the light absorption~\cite{fial1,fial2,fial3,Nosotros}, while the second part describes the term that breaks parity, which we are interested to explore in this work.

Following Refs.~\cite{fial1,fial2,fial3,Nosotros}, we interpret the delta function in Eq.~(\ref{maxwell}) as a current in the plane the graphene membrane. Thus, from Ohm's law, 
\begin{equation}
j_{a}=\sigma_{ab} E_{b}\;.
\label{ohm}
\end{equation} 
Assuming a varying incident electric wave with frequency $\omega$ expressed in a temporal gauge $A_0=0$, we can take $E_{b}=i\omega A_{b}$, which from the generalized Maxwell's equations~(\ref{maxwell}) allows us to write $j_{a}\simeq \Pi_{ab} A_{b}$. Finally, we can identify from Ohm's law, the conductivity tensor as
\begin{equation}
\sigma_{ab}=\frac{\Pi_{ab}}{i\omega}\;.
\label{transversesigma}
\end{equation}  
This expression becomes useful in what follows.

For the FRIM problem, let us consider a plane wave of frequency $\omega$, which travels along the $z$-direction from below the graphene layer with a linear polarization along the $\hat{e}_x$ direction. Moreover, considering that the wave passes through the graphene plane, the reflected and transmitted waves can be described as 
\begin{equation}
A=e^{-i\omega t}\left\{ \begin{array}{cc} \hat{e}_x e^{ik_z z}+(r_{xx}\hat{e}_x+r_{xy}\hat{e}_y)e^{-ik_z z}, & z<0,\\
(t_{xx}\hat{e}_x+t_{xy}\hat{e}_y)e^{ik_z z}, & z>0,\end{array}
\right.
\end{equation}
where $\hat{e}_{x,y}$ are the unit vectors along the directions $x$ and $y$ on the membrane.  The boundary conditions~(\ref{bc}) simplify to
\begin{eqnarray}
A_{a}\Bigg|_{z=0^+}-A_{a}\Bigg|_{z=0^-}&=&0\;,\nonumber\\
(\partial_z A_{a})\Bigg|_{z=0^+}-(\partial_z A_{a})\Bigg|_{z=0^-}&=&\alpha \left[ \Psi(\omega) \delta^{ab}+i\omega \Phi(\omega)\epsilon_a^{\ b} \right] A_{b} \Bigg|_{z=0}\;,
\label{bc2}
\end{eqnarray}
such that the transmission coefficients can be straightforwardly obtained~\cite{fial1,fial2,fial3}
\begin{eqnarray}
t_{xx}=\frac{-2\omega(i\alpha\Psi+2\omega)} {\alpha^2\Psi^2-4i\alpha\omega\Psi-(4+\alpha^2\Phi^2)\omega^2}\;,\nonumber\\
t_{xy}=\frac{2\alpha\Phi\omega^2} {\alpha^2\Psi^2-4i\alpha\omega\Psi-(4+\alpha^2\Phi^2)\omega^2}\;.
\end{eqnarray}
The angle of polarization rotation $\theta_F$ is obtained as follows:
\begin{eqnarray}
\theta_F&=&\frac{1}{2}\textrm{arg}\frac{t_{xx}-it_{xy}}{t_{xx}+it_{xy}}\nonumber\\
&=&-\frac{1}{2}\textrm{arg}\frac{i\alpha\Psi+2\omega+i\alpha\Phi\omega}{i\alpha\Psi+2\omega-i\alpha\Phi\omega}\nonumber\\
&=&-\frac{\alpha}{2}\textrm{Re}(\Phi)+{\cal O}(\alpha^2)\nonumber\\
&=&-\frac{1}{2}\frac{\Pi_{xy}}{i\omega}+{\cal O}(\alpha^2)\;.
\end{eqnarray}
Therefore, in terms of the Hall conductivity, the angle of FRIM $\theta_F$ becomes 
\begin{equation}
\theta_F=-\frac{\textrm{Re}(\sigma_{xy})}{2}+{\cal O}(\alpha^2)\;.
\end{equation}
Substituting the explicit form of $\sigma$ from the Eq.~(\ref{SigmaFilling}), we finally arrive at the main result of this article, namely,
\begin{eqnarray}
\theta_F=\left\{ \begin{array}{cc} 0, & m_0=0\\
\frac{e^2}{4\pi}\simeq\alpha, & m_e=0\end{array}\right. \;.
\end{eqnarray}
Observe that FRIM effect would appear only when the parity and time reversal symmetries are broken, in consistency with the experimental and theoretical findings for these quantities in absence of external fields as well as in and the presence of a strong magnetic field~\cite{fial1,faradayexp,kerr,aires}.

\section{Final remarks}

In this work, we have calculated the transverse (Hall) conductivity in a low energy effective model of graphene within a RQED$_3$ framework, in which we consider an ordinary mass term $m\psi\bar{\psi}$ that is invariant under $\mathcal{P}$ and $\mathcal{T}$ and the Haldane mass term $m_0\psi\tau\bar{\psi}$ that is invariant under pseudo-chiral transformations, Eq.~(\ref{chiral}), but breaks parity and time reversal. Those mass parameters might be thought as effective parametrizations of strains and other distortions in the crystalline structure of pristine graphene. Considering both the mass terms, the Kubo formula yields a half-filling factor per fermion species so long ${\cal P}$ and ${\cal T}$ symmetries are explicitly broken. 
In our case, we obtained $\nu = 1$. 
From the corresponding Hall conductivity,  we have estimated the Faraday rotation angle $\theta_F$ of an incident wave. Finally, we observe that the angle of polarization rotation even in the absence of external magnetic fields is $\theta_F=0$ when $m_0=0$ and $\theta_F=\alpha$ when $m_e=0$. This opens the possibility to measure that the polarization rotation with visible light in deformed graphene  and might be helpful to design optical filters. 

\section*{Acknowledgments}

We acknowledge M. Loewe for fruitful discussions and carefully reading of the manuscript. AR and SHO acknowledge CONACyT (M\'exico) and CIC-UMSNH  under grants 128534 and 4.22, respectively.  DV acknowledges support from CONICYT (Chile). SSM acknowledges CGIC-UCol (M\'exico).

\end{document}